\documentclass[12pt,a4paper]{article}

\usepackage{graphicx}

\input epsf

\newcommand{\frat}[2]{\frac{\textstyle #1}{\textstyle #2}}
\newcommand{\vf}[1]{\mbox{\boldmath $#1$}}

\begin{document}

\begin{center}
{\Large \bf  New arrangement of common approach to calculating the QCD ground state}\\
\vspace{0.3cm}
S. V. Molodtsov$^{1,2}$,  G. M. Zinovjev$^{3}$ \\
\vspace{0.3cm}
{\small
$^1$Joint Institute for Nuclear Research, RU-141980, Dubna, Moscow
region, RUSSIA.}\\
\vspace{0.3cm}
{\small $^2$Institute of
Theoretical and Experimental Physics, RU-117259, Moscow,
RUSSIA.}\\
\vspace{0.3cm}
{\small
$^3$Bogolyubov Institute for Theoretical Physics, UA-03143, Kiev,
UKRAINE.}
\end{center}
\vspace{0.3cm}

\begin{center}
\begin{tabular}{p{16cm}}
{\small{The quark behaviour in the background of intensive stochastic gluon field is
studied. An approximate procedure for calculating the effective Hamiltonian is
developed and the corresponding ground state within the Hartree-Fock-Bogolyubov
approach is found. The comparative analysis of various model Hamiltonian is given
and transition to the chiral limit in the Keldysh model is discused in detail.}}
\end{tabular}
\end{center}
\vspace{0.3cm}

We study the quark (anti-quark) behaviour while being influenced by intensive stochastic
gluon field and work in the context of the Euclidean field theory. The corresponding
Lagrangian density is the following
\begin{equation}
\label{1}
{\cal L}_E=\bar q~(i\gamma_\mu D_\mu+im)~q~,
\end{equation}
here  $q$ ($\bar q$) --- are the quark (anti-quarks) fields with covariant derivative
$D_\mu=\partial_\mu -i g A^a_\mu t^a$ where $A^a_\mu$ is the  gluon field, $t^a=\lambda^a/2$
are the generators of colour gauge group $SU(N_c)$ and $m$ is the current quark mass.
As the model of stochastic gluon field we refer to the example of (anti-)instantons considering
an ensemble of these quasi-classical configurations.
On our way to construct an effective theory
(which usually encodes the predictions of a quantum field theory at low energies) the assumptions
done are not of special importance. However, what is entirely restrictive to fix the effective
action at really low energy (i.e. low cutoff) up to a few coupling constants is an idea to neglect
all the contributions coming from gluon fields $A_{ex}$ generated by the (anti-)quarks.
$$A_{ex} \ll A~.$$
Actually, it means the removal of corresponding cutoff(s) from consideration,
but by the definition
of an effective theory this operation does not pose itself.
Then the corresponding Hamiltonian description results from
\begin{equation}
\label{2}
{\cal H}=\pi \dot q-{\cal L}_E~,~~\pi=\frat{\partial{\cal L}_E}{\partial \dot q}=i q^{+}~,
\end{equation}
and
\begin{equation}
\label{3}
{\cal H}_0=-\bar q~(i{\vf \gamma}{\vf \nabla}+im)~q~,
\end{equation}
for noninteracting quarks.
In Schr\"odinger representation the quark field evolution is determined by the equation for
the quark probability amplitude $\Psi$ as
\begin{equation}
\label{4}
\dot \Psi=- H \Psi~,
\end{equation}
with the density of interaction Hamiltonian
\begin{equation}
\label{6}
{\cal V}_S=\bar q({\vf x})~t^a\gamma_\mu A^{a}_\mu(t,{\vf x})~q({\vf x})~.
\end{equation}
The explicit dependence on "time" is present at the gluon field only. The creation and annihilation
operators of quarks and anti-quarks $a^+, a$, $b^+, b$ have no "time" dependence and consequently
\begin{equation}
\label{5}
q_{\alpha i}({\vf x})=\int\frat{d {\vf p}}{(2\pi)^3} \frat{1}{(2|p_4|)^{1/2}}~
\left[~a({\vf p},s,c)~u_{\alpha i}({\vf p},s,c)~ e^{i{\vf p}{\vf x}}+
b^+({\vf p},s,c)~v_{\alpha i}({\vf p},s,c)~ e^{-i{\vf p}{\vf x}}\right]~.
\end{equation}
The stochastic character of gluon field (which we supposed) allows us to develop the approximate
description of the state $\Psi$ if the following procedure of averaging
$$\Psi \to \langle \Psi \rangle=\int_0^t ~d\tau ~\Psi(\tau)/~t$$
is intoduced. With this procedure taken the futher step is to turn to the approach of constructing
a density matrix $\langle \stackrel{*}{\Psi}\stackrel{}{\Psi} \rangle$. However, here we believe
that at calculating the ground state (or more generally with quasi-stationary state) it might be
sufficiently informative to operate with the averaged amplitude directly. Then in the interaction
representation  $\Psi=e^{H_0 t}\Phi$ we have the equation for state $\Phi$ as
\begin{equation}
\label{7}
\dot \Phi=- V \Phi~,~~V=e^{H_0 t} V_S e^{-H_0 t}~.
\end{equation}
Now the "time" dependence appears in quark operators as well and after averaging over the
short-wavelength component one may obtain the following equation
\begin{equation}
\label{8}
\langle\dot\Phi(t)\rangle=+\int_0^{\infty} d\tau ~\langle V(t) V(t-\tau)\rangle~\langle\Phi(t)\rangle~.
\end{equation}
The limitations to have such a factorization validated are well known in the theory of stochastic
differential equations (see, for example, \cite{van}). The integration interval in Eq.(8) may be
extended to the infinite "time" because of the rapid decrease (supposed) of the corresponding
correlation function. Now we are allowed to deal with amplitude $\langle\Phi(t)\rangle$ in the right
hand side of Eq.(8) instead the amplitude with the shifted arguments in order to get an ordinary
integro-differential equation. In the quantum field theory applications it is usually difficult to
construct the correlation function in the most general form. However, if we are going to limit our
interest by describing the long-wavelength quark component only then gluon field correlator
$\langle A^a_\mu(x) A^b_\nu(y)\rangle$ may be factorized and as a result we have
$$
\langle\dot\Phi(t)\rangle=\int d{\vf x}~ \bar q({\vf x},t)~t^a\gamma_\mu~q({\vf x},t)~
\int_0^{\infty} d\tau \int d{\vf y} ~\bar q({\vf y},t-\tau)~t^b\gamma_\nu~q({\vf y},t-\tau)~
g^2\langle A^{a}_\mu(t,{\vf x}) A^{b}_\nu(t-\tau,{\vf y})\rangle~
\langle\Phi(t)\rangle~.$$
Having assumed the correlation function rapidly decreasing in "time" we could ignore all the
retarding effects in the quark operators.
Turning back to the Schr\"odinger representation we have for the state amplitude
$\chi=e^{-H_0 t}\langle\Phi\rangle$ the following equation
\begin{eqnarray}
\label{9}
&&\dot \chi=- H_{ind}~ \chi~,\nonumber\\ [-.2cm]
\\ [-.25cm]\nonumber
&&{\cal H}_{ind}=-\bar q~(i{\vf \gamma}{\vf \nabla}+im)~q-\bar q~t^a\gamma_\mu~q~
  \int d{\vf y} ~\bar q'~t^b\gamma_\nu~q'~
\int_0^{\infty} d\tau~ g^2\langle A^{a}_\mu A^{'b}_\nu\rangle~,\nonumber
\end{eqnarray}
with $q=q({\vf x})$, $\bar q=\bar q({\vf x})$, $q'=q({\vf y})$, $\bar q'=\bar q({\vf y})$ and
$A^{a}_\mu =A^{a}_\mu(t,{\vf x})$, $A^{'b}_\nu=A^{b}_\nu(t-\tau,{\vf y})$. Now the correlation
function might be presented as
$$\int_0^{\infty} d\tau~g^2\langle A^{a}_\mu A^{'b}_\nu\rangle=
\delta^{a b}~F_{\mu\nu}({\vf x}-{\vf y})~,$$
with the corresponding formfactors
$F_{\mu\nu}({\vf x}-{\vf y})=\delta_{\mu\nu}~I({\vf x}-{\vf y})+J_{\mu\nu}({\vf x}-{\vf y})$.
In our consideration we ignore the contribution of the second formfactor spanning on the components
of the vector ${\vf x}-{\vf y}$. Thus, on output we receive the Hamiltonian
of four-fermion interaction with the formfactor rooted in the presence of two quark currents
in the points ${\vf x}$ and ${\vf y}$.
With this form of the effective Hamiltonian we could apply the Hartree--Fock--Bogolyubov
method \cite{nnb} to find its ground state as one constructed by the quark--anti-quark pairs
with the oppositely directed momenta
\begin{eqnarray}
\label{10}
&&|\sigma\rangle=T~|0\rangle~,\nonumber\\ [-.2cm]
\\ [-.25cm]
&&
T=\Pi_{ p,s,c}~\exp\left\{~\frac{\theta}{2}~\left[~a^+({\vf p},s,c)~b^+(-{\vf p},s,c)+
a({\vf p},s,c)~b(-{\vf p},s,c)~\right]~\right\}~,\nonumber
\end{eqnarray}
where the parameter $\theta({\vf p})$ characterizes the pairing strength. Introducing the
creation and annihilation operators of quasi-particles
$A=T~a~T^{-1}$, $A^+=T~a^+T^{-1}$, $B=T~b~T^{-1}$, $B^+=T~b^+T^{-1}$,
we can rewrite the quark (anti-quark) operators as
\begin{eqnarray}
\label{12}
&&q({\vf x})=\int\frat{d {\vf p}}{(2\pi)^3} \frat{1}{(2|p_4|)^{1/2}}~
\left[~A({\vf p},s,c)~U({\vf p},s,c)~e^{i{\vf p}{\vf x}}+
B^+({\vf p},s,c)~V({\vf p},s,c)~ e^{-i{\vf p}{\vf x}}\right]~,\nonumber\\
&&\bar q({\vf x})=\int\frat{d {\vf p}}{(2\pi)^3} \frat{1}{(2|p_4|)^{1/2}}~
\left[~A^+({\vf p},s,c)~\overline{U}({\vf p},s,c)~e^{-i{\vf p}{\vf x}}+
B({\vf p},s,c)~\overline{V}({\vf p},s,c)~ e^{i{\vf p}{\vf x}}\right]~,\nonumber
\end{eqnarray}
with the quasi-particle spinors
\begin{eqnarray}
\label{13}
&&U({\vf p},s,c)=\cos\left(\frac{\theta}{2}\right)~u({\vf p},s,c)-
\sin\left(\frac{\theta}{2}\right)~v(-{\vf p},s,c)~,\nonumber\\ [-.2cm]
\\ [-.25cm]
&&V({\vf p},s,c)=\sin\left(\frac{\theta}{2}\right)~u(-{\vf p},s,c)+
\cos\left(\frac{\theta}{2}\right)~v({\vf p},s,c)~,\nonumber
\end{eqnarray}
where $\overline{U}({\vf p},s,c)=U^+({\vf p},s,c)~\gamma_4$,
$\overline{V}({\vf p},s,c)=V^+({\vf p},s,c)~\gamma_4$. Minimizing the mean energy
functional one is able to determine the angle $\theta$ magnitude
\begin{equation}
\label{27}
\frat{d\langle\sigma|H_{ind}|\sigma\rangle}{d\theta}=0~.
\end{equation}
Dropping the calculation details out we present here the following result for the mean energy
as a function of the $\theta$ angle
\begin{eqnarray}
\label{26}
&&\langle\sigma|H_{ind}|\sigma\rangle=-\int \frat{d {\vf p}}{(2\pi)^3}~\frat{2N_c~p_4^{2}}{|p_4|}
\left(1-\cos\theta\right)-
\nonumber\\
&&-\widetilde G\int \frat{d {\vf p}d {\vf q}}{(2\pi)^6}\left\{-(3\widetilde I-\widetilde J)
\frat{p_4~ q_4}{|p_4||q_4|}+(4\widetilde I-\widetilde J)\frat{p~ q}{|p_4||q_4|}
\left(\sin\theta-\frat{m}{p}\cos\theta\right)
\left(\sin\theta'-\frat{m}{q}\cos\theta'\right)+\right.\nonumber\\ [-.2cm]
\\ [-.25cm]
&&
\left.+(-2\widetilde I\delta_{ij}-2\widetilde J_{ij}+\widetilde J\delta_{ij})~\frat{p_i~q_j}{|p_4||q_4|}~
\left(\cos\theta+\frat{m}{p}\sin\theta\right)
\left(\cos\theta'+\frat{m}{ q}\sin\theta'\right)~\right\}~,\nonumber
\end{eqnarray}
here the following designations are used $p=|{\vf p}|$, $q=|{\vf q}|$,
$\widetilde I=\widetilde I({\vf p}+{\vf q})$, $\widetilde J_{ij}=\widetilde J_{ij}({\vf p}+{\vf q})$,
$\widetilde J=\sum_{i=1}^3\widetilde J_{ii}$,
$p^2=q^2=-m^2$, $\theta'=\theta(q)$ where $\widetilde G$ is the constant of corresponding
four-fermion interaction (the relevant details can be found in \cite{MZ}).
The first integral in Eq. (\ref{26}) comes from free Hamiltonian, and we make a natural subtraction (adding the unit)
in order to have zero mean free energy when the angle of pairing is trivial.
\\

\begin{figure*}[!tbh]
\begin{center}
\includegraphics[width=0.45\textwidth]{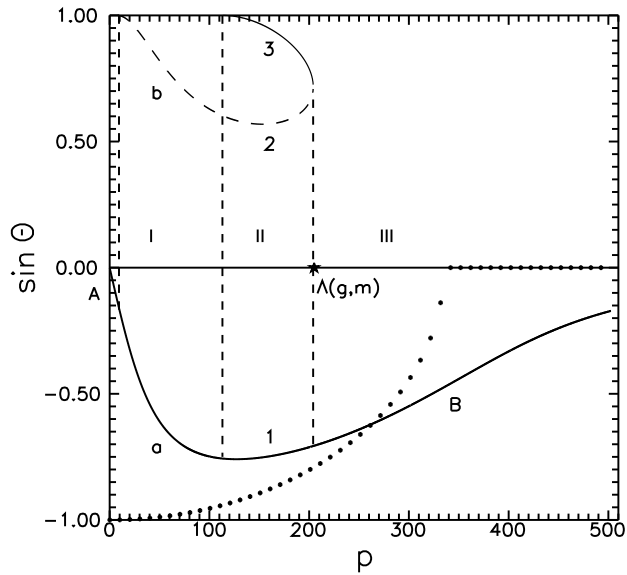}
\end{center}
  \vspace{-7mm}
 \caption{Phase portrait of the Keldysh model, $\sin\theta$ as a function of momentum $p$(MeV). The dotted
curve  corresponds to the solution with the negative values of angle in the chiral limit $m=0$.}
  \label{f1}
\end{figure*}

{\bf Nambu--Jona-Lasinio model}\\
\\
In order to get an idea of the parameter scales we continue with handling the model in which the formfactor
behaves in the coordinate space as
$I({\vf x}-{\vf y})=\delta({\vf x}-{\vf y})$, $J_{\mu\nu}=0$, dropping contribution
spanned on the $p_i q_j$ tensor also. Actually, it corresponds to
the Nambu--Jona-Lasinio model \cite{njl}. As well known the model with such a formfactor
requires the regularization and, hence, the cutoff parameter $\Lambda$ comes to the play
\begin{equation}
\label{28}
W=\int^{\Lambda}\frat{d{\vf p}}{(2\pi)^3}~\left[|p_4|\left(1-\cos\theta\right)-
G\frat{p}{|p_4|}\left(\sin\theta-\frat{m}{p}\cos\theta\right)\int^\Lambda \frat{d{\vf q}}{(2\pi)^3}
\frat{q}{|q_4|}\left(\sin\theta'-\frat{m}{q}\cos\theta'\right)\right]~.
\end{equation}
We adjust the NJL model with the parameter set given by Hatsuda and Kunihiro \cite{njl} in which
$\Lambda=631 \mbox{MeV}$, $m=5.5\mbox{MeV}$. One curious point of this model is that the solution
for optimal angle $\theta$ in the whole interval $p\in[0,\Lambda]$ can be found by solving
the simple trigonometrical equation
\begin{equation}
\label{29}
(p^2+m^2)~\sin\theta-M_q\left(p\cos\theta+m\sin\theta\right)=0~,
\end{equation}
with the dynamical quark mass
\begin{equation}
\label{30}
M_q=2G~\int^\Lambda \frat{d{\vf p}}{(2\pi)^3}\frat{p}{|p_4|}
~\left(\sin\theta-\frat{m}{p}\cos\theta\right)~.
\end{equation}
Eventually the results obtained look like $M_q=-335$ MeV for dynamical quark mass and
$\langle\sigma|\bar q q|\sigma\rangle=-i~(245$ MeV$)^3$ for the quark condensate with the following
definition of the quark condensate
\begin{equation}
\label{37}
\langle\sigma|\bar q q|\sigma\rangle=
\frat{i~N_c}{\pi^2}~\int_0^\infty dp~\frat{p^2}{|p_4|}~(p\sin\theta-m\cos\theta)~.
\end{equation}

{\bf The Keldysh model}\\
\\
Now we are going to analyse the limit, in some extent, opposite to the NJL model, i.e. we are
dealing with the formfactor behaving as a delta function but in the momentum space
(analogously the Keldysh model, well known in the physics of condensed matter \cite{K}),
$I({\vf p})=(2\pi)^3~\delta({\vf p})$. Here the mean energy functional has the following form
\begin{equation}
\label{31}
W(m)=\int \frat{d{\vf p}}{(2\pi)^3}~\left[|p_4|~\left(1-\cos\theta\right)-
G~\frat{p^2}{|p_4|^2}\left(\sin\theta-\frat{m}{p}\cos\theta\right)^2\right]~.
\end{equation}
contrary to the NJL model there is no need to introduce any cut off.
The equation for calculating the optimal angle $\theta$ becomes the transcendental one
\begin{equation}
\label{32}
|p_4|^3~\sin\theta-2G~\left(p\cos\theta+m\sin\theta\right)
\left(p\sin\theta-m\cos\theta\right)=0~,
\end{equation}
and, clearly, it is rather difficult to get its solution in a general form. Fortunately, it is
much easier and quite informative to analyse the model in the chiral limit $m=0$.
There  exist one trivial solution $\theta=0$ and two nontrivial ones (for the positive and
negative angles) which obey the equation
\begin{equation}
\label{33}
\cos\theta=\frat{p}{2G}~.
\end{equation}
Obviously, these solutions are reasonable if the momentum is limited by $p<2G$. Then for
the mean energy we have
$W_\pm(0)=-\frat{G^4}{15\pi^2}$ if the quark condensate defined as
$\langle\sigma|\bar q q|\sigma\rangle(0)=\frat{i~N_c~G^3}{2\pi}$.
For the trivial solution the mean energy equals to zero together with the quark condensate
$W_0(0)=0$, $\langle\sigma|\bar q q|\sigma\rangle_0(0)=0$. Introducing the practical designation
$\sin\theta=\frat{M_\theta}{(p^2+M^2_\theta)^{1/2}}$ which characterizes the pairing strength
by the parameter $M_\theta$ we have, for example, for the nontrivial solution
$M_\theta=\left(4G^2-p^2\right)^{1/2}$. In order to compare the results with the NJL model
we fixed the value of four-fermion interaction constant as $M_\theta(0)=2G=335$ MeV.
It is interesting to notice that the respective energy becomes constant
$E(p)=\sqrt{p^2+M_\theta^2}$, $E(p)=2G$.

\begin{figure*}[!tbh]
\begin{center}
\includegraphics[width=0.45\textwidth]{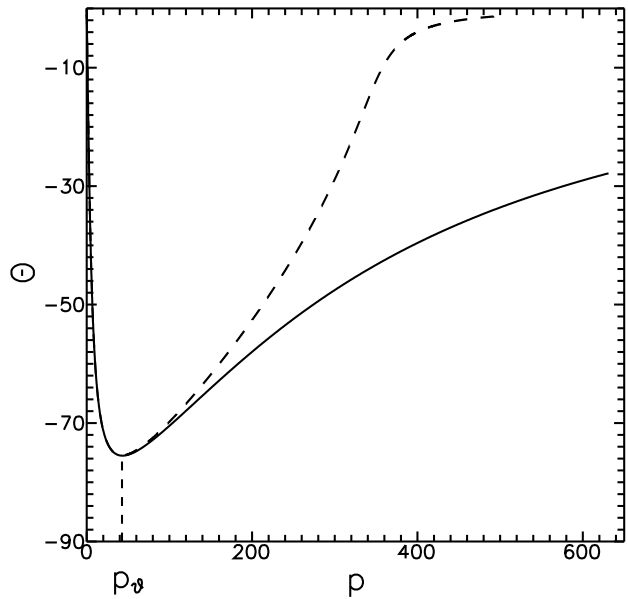}
\end{center}
  \vspace{-7mm}
 \caption{The optimal angle $\theta$ as a function of momentum $p$(MeV). The solid line corresponds
to the NJL model and the dashed one to the Keldysh model. The current quark mass is $m=5.5$ MeV and
$p_\theta\sim 40$ MeV.}
  \label{f4}
\end{figure*}

After having done the analysis in the chiral limit which is shown by the dotted line in Fig.1 we
would like to comment the situation beyond this limit. The evolution of corresponding branches is
available on the same plot \ref{f1}. One solution denoted by A is developing in the local vicinity
of coordinate origin and for small values of quark mass this domain is practically indistinguishable.
In order to make it noticeable (to have a reasonable resolution on the plot) the quark mass was
put as $m=50$ MeV. Besides, there are two solutions $a$ and $b$ in the domain denoted by $I$, three
solutions denoted by $1$, $2$, $3$ in the domain $II$ and one solution $B$ in the domain $III$. The
minimum of mean energy functional can be realized with the piecewise continuous functions.
At the local vicinity of coordinate origin we start with the solution branch A, then relevant solution
passes to the branch $a$ or $b$ interchanging its position from $a$ to $b$ in any subinterval. But in any case
there is only one way to continue the solution at streaming to the infinite limit and it is related
with the branch $B$ where the angle is going to the zero value. As to the functional (\ref{31}) the
contribution of the term proportional to the cosine in the second parenthesis is divergent even if the
angle $\theta$ is zero. It means the mean energy out of chiral limit goes to an infinity at any
nonzero value of quark mass. The same conclusion is valid for
the chiral condensate. In principle this functional could be regularized and corresponding continuation
might be done but it is out of this presentation scope. It is not difficult to demonstrate the
similar discontinuities of functional are present, for example, for Gaussian
$
I({\vf x})=G~\exp{(-a^2~{\vf x}^2)}$,
and exponential
$
I({\vf x})=G~\exp{(-a~|{\vf x}|)}$,
formfactors and they are present even in the NJL model but this fact is masked by the cut off parameter.

\begin{figure*}[!tbh]
\begin{center}
\includegraphics[width=0.45\textwidth]{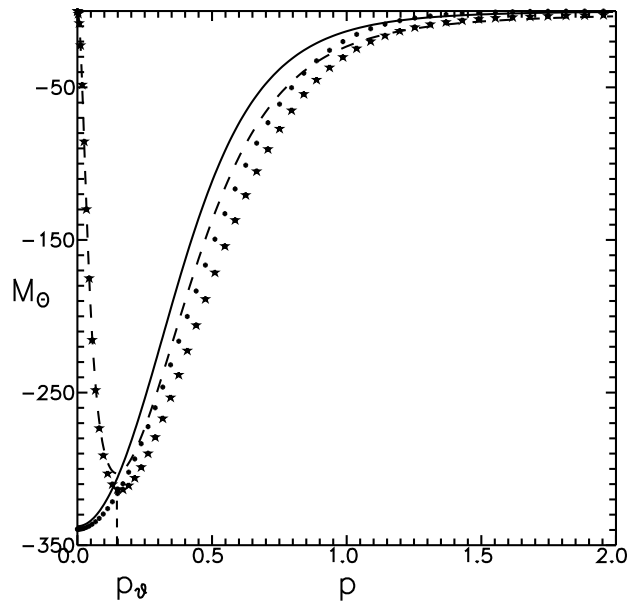}
\end{center}
  \vspace{-7mm}
 \caption{The parameter $M_\theta$(MeV) as a function of momentum $p$(GeV) corresponding to
the best fit of the NJL data $M_q=335$ MeV, $\langle\sigma|\bar q q|\sigma\rangle=-i~(245$ MeV$)^3$.
The solid line corresponds to the Gaussian formfactor in chiral limit and the dashed line
corresponds to the magnitude of current quark mass $m=5.5$ MeV.
The exponentially behaving formfactor is represented by the dotted lines and $p_\theta\sim 150$ MeV.}
  \label{f7}
\end{figure*}

Comparing the optimal angles in the NJL and Keldysh models (see Fig. 2) it is interesting to notice that the
formation of quasiparticles becomes significant at some momentum value close to the origin
$p_\theta\sim 40$ MeV but not directly at the zero value. It is clear the inverse value of this parameter
determines the characteristic size of quasiparticle. Parameter $M_\theta$ as a function of momentum $p$
corresponding to the best fit to the NJL data $M_q=335$ MeV,
$\langle\sigma|\bar q q|\sigma\rangle=-i~(245$ MeV$)^3$ is shown in Fig.3. The solid line corresponds to
the Gaussian formfactor in the chiral limit and the dashed one shows the same dependence for the
current quark mass $m=5.5$ MeV. This dependence for exponential behaviour of formfactor is
presented by the dotted lines on the same plot (the characteristic angle is $p_\theta\sim 150$ MeV in this
case).
Analysing the discontinuity of mean energy functional and quark condensate we face some
troubles at fitting the quark condensate, for example. However, the dynamical quark mass and quark
condensate are nonobservable quantities and it is curious to remark here that although the mean
energy of the quark system is minus infinity the meson observables are finite and even
in Keldysh model the mesons are recognizable with reasonable scale
and we can in principle make a fit for this observables.

\end{document}